\documentstyle[12pt,epsf]{article}
\def\ga{\mathrel{\raise.3ex\hbox{$>$\kern-.75em\lower1ex\hbox{$\sim$}}}}
\def\la{\mathrel{\raise.3ex\hbox{$<$\kern-.75em\lower1ex\hbox{$\sim$}}}}
\def\he#1{\hbox{${}^{#1}$He}}
\def\li#1{\hbox{${}^{#1}$Li}}
\def\beq{\begin{equation}}
\def\eeq{\end{equation}}

\begin{document}
\pagestyle{empty}
\baselineskip=13pt
\rightline{UMN--TH--1419/95}
\rightline{hep-ph/9512166}
\rightline{December 1995}
\vspace*{3.2cm}
\begin{center}
{\large{
BIG BANG NUCLEOSYNTHESIS AND THE CONSISTENCY BETWEEN THEORY
AND THE OBSERVATIONS OF \newline
D, \he3, \he4, AND \li7 }}
\end{center}
{}~\newline

\baselineskip=2ex
\begin{center}
{\large Keith A.~Olive
}\\
{\large \it
{School of Physics and Astronomy,
University of Minnesota,\\ Minneapolis, MN 55455, USA}}

\vspace*{3cm}
{\bf Abstract}
\end{center}
The current status of big bang nucleosynthesis is reviewed.
 Particular attention is given to the degree at which the theory is
consistent with the observation of the light element abundances.

\vspace*{4.5ex}
\baselineskip=3ex
The observational information at hand on the abundances of the four light
element isotopes: D, \he3, \he4, and \li7 can be characterized as
being either reasonably certain (for \he4 and \li7); getting certain (for
D); or uncertain (for \he3). In the cases of \he4 and \li7, there is a
growing wealth of data from extragalactic HII regions$^{1)}$ on \he4 and
from the surfaces of old Population II halo stars$^{2)}$ on \li7. Indeed, we
are
rapidly approaching the point where our uncertainty in the abundances of these
isotopes is dominated by systematic rather than statistical uncertainties.
For D, though we have good solar and interstellar medium (ISM) data$^{3)}$,
the connection to a primordial abundance through galactic chemical
evolution introduces
substantial uncertainties.  However, recent observations of D in quasar
absorption systems are beginning to to yield a more coherent picture for
what may be the true primordial abundance of D.
In the uncertain category, I would place \he3.  As for D, we have data
in the ISM and there are also \he3 abundances in planetary nebulae, however
in this case, not only our we hampered by our lack of knowledge concerning
galactic chemical evolution, but also by our uncertainties in the production
of \he3 in low mass stars.

Overall, there is certainly a broad agreement between the predicted abundances
of the light element isotopes from big bang nucleosynthesis$^{4)}$
and the abundances
inferred from observations which span nearly ten orders of magnitude.
Indeed, the standard model theory really only contains one free parameter,
namely, the baryon to photon ratio, $\eta$. Agreement between theory and
the observation of all of the four isotopes occurs when
$\eta_{10} = 10^{10}\eta$ is in the range,  2.8 -- 4. However, at  closer
level, the above range on $\eta$ is slightly high with regard to what is
preferred by \he4, and may be too low for D and \he3 when chemical
evolution is combined with the observations. In fact, the lower bound to $\eta$
comes directly from an upper bound to the combined abundance of D and \he3 and
depends on the degree to which \he3 survives in stellar evolution.
Recent data on D and \he3 may be yielding an inconsistent picture. As argued
below, I believe that the most likely source of the problem is our treatment
of \he3 due to the considerable uncertainties in chemical and stellar evolution
concerning this isotope. Furthermore, I will show that  \he4 and \li7
(abundances that we know best) are already sufficient in constraining the
theory
and that the abundance of D/H predicted from \he4 and \li7, appears to be
consistent with the recent measurements of D/H in quasar absorption systems,
though this requires that some of our beliefs concerning \he3 need rethinking.

Before moving to the comparison of theory and observation, it will be
useful to first briefly review the observational status of the four
isotopes considered.  More detail on each can be found in refs. 1--3).

There is now a good collection of abundance information on the \he4 mass
fraction, $Y$, O/H, and N/H in over 50 extragalactic HII regions$^{5-7)}$.
The observation of the heavy elements is important as the helium mass fraction
observed in these HII regions has been augmented by some stellar processing,
the degree to which is given by the oxygen and nitrogen abundances.
In an extensive study based on the data in refs. 5) and 6), it was found$^{8)}$
that the data is well represented by a linear correlation for
$Y$ vs. O/H and Y vs. N/H. It is then expected that the primordial abundance
of \he4 can be determined from the intercept of that relation.  The overall
result
of that analysis indicated a primordial mass fraction,  $Y_p  = 0.232 \pm
0.003$.
In ref. 9), the stability of this fit was verified by a Monte-Carlo analysis
showing that the fits were not overly sensitive to any particular HII region.
In addition, the data from ref. 7) was also included, yielding a \he4 mass
fraction$^{9)}$
\beq
Y_p = 0.234 \pm 0.003 \pm 0.005
\label{he4}
\eeq
The second uncertainty is an estimate of the systematic uncertainty in the
abundance determination. Though the systematic uncertainty may be somewhat
larger$^{10)}$, its precise value will be superfluous to the discussion below.

As I have said above, I also believe that the \li7 abundance
is reasonably well known.
 In old,
hot, population-II stars, \li7 is found to have a very
nearly  uniform abundance. For
stars with a surface temperature $T > 5500$~K
and a metallicity less than about
1/20th solar (so that effects such as stellar convection may not be important),
the  abundances show little or no dispersion beyond that which is
consistent with the errors of individual measurements.
Indeed, as detailed by Spite$^{2)}$, much of the work concerning
\li7 has to do with the presence or absence of dispersion and whether
or not there is in fact some tiny slope to a [Li] = $\log$ \li7/H + 12 vs.
T or [Li] vs. [Fe/H] relationship.
There is \li7 data from nearly 100 halo stars, from a
 variety of sources. I will use the value given in ref. 11) as the best
estimate
for the mean \li7 abundance and its statistical uncertainty in halo stars
\beq
{\rm Li/H = (1.6 \pm 0.1 {}^{+0.4}_{-0.3} {}^{+1.6}_{-0.5}) \times 10^{-10}}
\label{li}
\eeq
 The first error is statistical, and the second
is a systematic uncertainty that covers the range of abundances
derived by various methods.
The third set of errors in Eq. (\ref{li}) accounts for
 the possibility that as much as half
of the primordial \li7 has been
destroyed in stars, and that as much as 30\% of the observed \li7 may have been
produced in cosmic ray collisions rather than in the Big Bang.
 Observations of \li6,
Be, and B help constrain the degree to which these effects
play a role$^{12)}$. For \li7, the uncertainties are clearly dominated by
systematic effects.

Turning, to D/H, we have three basic types of abundance information:
1) ISM data; 2) solar system information; and perhaps 3) a primordial
abundance from quasar absorption systems.  The best measurement for ISM D/H
is$^{13)}$
\beq
{\rm (D/H)_{ISM}} = 1.60\pm0.09{}^{+0.05}_{-0.10} \times 10^{-5}
\eeq
However, it is becoming apparent that this value may not be universal
(or galactic as the case may be) and that there may be some real dispersion
of D/H in the ISM$^{3)}$. The solar abundance of D/H is inferred from two
distinct measurements of \he3. The solar wind measurements of \he3 as well as
the low temperature components of step-wise heating measurements of \he3 in
meteorites yield the presolar (D + \he3)/H ratio as D was efficiently burned to
\he3 in the Sun's pre-main-sequence phase.  These measurements
indicate that$^{14,15)}$
\beq
{\rm \left({D +~^3He \over H} \right)_\odot = (4.1 \pm 0.6 \pm 1.4) \times
10^{-5}}
\eeq
 The high temperature components in meteorites are believed to yield the true
solar \he3/H ratio of$^{14,15)}$
\beq
{\rm \left({~^3He \over H} \right)_\odot = (1.5 \pm 0.2 \pm 0.3) \times
10^{-5}}
\label{he3}
\eeq
The difference between these two abundances reveals the presolar D/H ratio,
giving,
\beq
{\rm (D/H)_{\odot}} \approx (2.6 \pm 0.6 \pm 1.4) \times 10^{-5}
\eeq
Finally, there are the recent measurements of D/H in quasar absorption systems.
The first of these measurements$^{16)}$ indicated a rather high D/H ratio,
D/H $\approx$ 1.9 -- 2.5 $\times 10^{-4}$. However, a reported
measurement$^{17)}$ of  D/H in a second system seemed to show a very different
abundance, D/H $\approx$ 1 -- 2 $\times 10^{-5}$. Most recently, a new
observation$^{18)}$ of  the high D/H absorber was made resolving it into two
components. The weighted average of these two components
indicates that D/H = $(1.9 \pm 0.4) \times 10^{-4}$ in these systems, again
calling for a high primordial D/H. It is probably premature to use this value
as
the primordial D/H abundance in an analysis of big bang nucleosynthesis, but it
is certainly encouraging that future observations may soon yield a firm
value for D/H. It is however important to note that there does seem to be a
trend that over the history of the Galaxy, the D/H ratio  is decreasing,
something we expect from galactic chemical evolution.  Of course the total
amount of deuterium astration that has occurred is still uncertain, and model
dependent.

There are also several types of \he3 measurements. As noted above, meteoritic
extractions yield a presolar value for \he3/H as given in Eq. (\ref{he3}).
In addition, there are several ISM measurements of \he3 in galactic HII
regions$^{19)}$ which also show a wide dispersion
\beq
 {\rm \left({~^3He \over H} \right)_{ISM}} \simeq 1 - 5 \times 10^{-5}
\eeq
  Finally there are observations of \he3 in planetary
nebulae$^{20)}$ which show a very high \he3 abundance of \he3/H $\sim 10^{-3}$.

Each of the light element isotopes can be made consistent with theory for a
specific range in $\eta$. Overall consistency of course requires that
the range in $\eta$ agree among all four light elements.
\he3 (together with D) has stood out in its importance for BBN, because
it  provided a (relatively large) lower limit for the baryon-to-photon
ratio$^{21)}$, $\eta_{10} > 2.8$. This limit for a long time was seen to be
essential because it provided the only means for bounding $\eta$ from below
and in effect allows one to set an upper limit on the number of neutrino
flavors$^{22)}$, $N_\nu$, as well as other constraints on particle physics
properties. That is, the upper bound to $N_\nu$
is strongly dependent on the lower bound to
$\eta$.  This is easy to see: for lower $\eta$, the \he4 abundance drops,
allowing for a larger $N_\nu$, which would raise the \he4 abundance.
However, for $\eta < 4 \times 10^{-11}$, corresponding to $\Omega h^2 \sim
.001-.002$ which is not too different from galactic mass densities, there is no
bound whatsoever on $N_\nu$$^{23)}$. Of course, with the improved data on \li7,
we
do have lower bounds on $\eta$ which exceed $10^{-10}$.

 In ref. 21), it was argued that since stars (even massive stars) do not
destroy \he3 in its entirety, we can obtain a bound on $\eta$ from an
upper bound to the solar D and \he3 abundances. One can in fact
limit$^{21,24)}$
 the sum of primordial D and \he3 by applying the expression below at $t =
\odot$
\beq
{\rm \left({D + \he3 \over H} \right)_p \le \left({D \over H} \right)_t}
+ {1 \over g_3}{\rm  \left({\he3 \over H} \right)_t}
\label{he3lim}
\eeq
In (\ref{he3lim}), $g_3$ is the fraction of a star's initial D and \he3 which
survives as \he3. For $g_3 > 0.25$ as suggested by stellar models, and using
the
solar data on D/H and
\he3/H, one finds $\eta_{10} > 2.8$. This argument has been improved
recently$^{25)}$ ultimately leading to a stronger limit$^{26)}$  $\eta_{10} >
3.8$
and a best estimate $\eta_{10} = 6.6 \pm 1.4$. The problem with this bound, is
that it seems to indicate an inconsistency most notably in the high \he4 mass
fraction predicted at the large value of $\eta$. It has been speculated that
the
cause may be underestimated systematic uncertainties in the \he4 abundance, a
problem with chemical evolution, or even a tau-neutrino mass thereby lowering
$N_\nu$ from  3 to 2. Indeed, at the large value of $\eta$, if $N_\nu $ is
allowed to be adjusted, a value around 2 is needed to match the \he4
mass fraction of 0.234. Even at $\eta$, around $\eta_{10} = 3$, the
preferred value for $N_\nu$ would be well below 3$^{8,27)}$.

The limit $\eta_{10} > 2.8 (3.8)$ derived using (\ref{he3lim}) is really a one
shot approximation.  Namely, it is assumed that material passes through a star
no
more than once. ( Although even the stochastic approach used in ref. 28) could
only lower the bound from 3.8 to about 3.5 when assuming as always that $g_3 >
0.25$). To determine whether or not the solar (and present) values of D/H and
\he3/H can be matched it is necessary to consider models of galactic chemical
evolution$^{29)}$. In the absence of stellar \he3 production, particularly by
low mass stars, it was shown$^{30)}$ that there are indeed suitable choices
for a star formation rate, and an initial mass function, to: 1) match the D/H
evolution from a primordial value (D/H)$_p = 7.5 \times 10^{-5}$,
corresponding to $\eta_{10} = 3$, through the solar and ISM abundances; while
2)
at the same time keeping the \he3/H evolution relatively flat so as not to
overproduce \he3 at the solar and present epochs. This was achieved for $g_3 <
0.3$. Even for $g_3 \sim 0.7$, the present \he3/H could be matched, though the
solar value was found to be a factor of 2 too high. For (D/H)$_p \simeq 2
\times
10^{-4}$, corresponding to  $\eta_{10} \simeq 1.7$, though models could be
found which destroy D sufficiently, overproduction of \he3 occurred unless
$g_3$ was tuned down to about 0.1.

In the context of models of galactic chemical evolution, there is however,
 little justification a
priori, for neglecting the production of \he3 in low mass
stars. Indeed, stellar models predict that considerable amounts of \he3 are
produced in stars between 1 and 3 M$_\odot$. For M $<$ 8M$_\odot$, Iben and
Truran$^{31)}$ calculate
\beq
(^3{\rm He/H})_f = 1.8 \times 10^{-4}\left({M_\odot \over M}\right)^2
+ 0.7\left[({\rm D+~^3He)/H}\right]_i
\label{it}
\eeq
so that at $\eta_{10} = 3$, and ((D + \he3)/H)$_i = 9 \times 10^{-5}$,
$g_3(1 $M$_\odot$) = 2.7! It should be emphasized that this prediction is in
fact consistent with the observation of high \he3/H in planetary
nebulae$^{20)}$.

Generally, implementation of the \he3 yield in Eq. (\ref{it}) in chemical
evolution models, leads to an overproduction of \he3/H particularly at the
solar epoch$^{32,33)}$. In Figure 1, the evolution of D/H and \he3/H is shown
as
a function of time from refs. 14,32). The solid curves show the evolution in a
 simple model of galactic chemical evolution with a star formation rate
proportional to the gas density and a power law IMF (see ref. 32) for details).
The model was chosen to fit the observed deuterium abundances. However, as one
can plainly see, \he3 is grossly overproduced (the deuterium data is
represented
by squares and \he3 by circles). Depending on the particular model chosen,
it may be possible to come close to at least the upper end of the range of
the \he3/H observed in galactic HII regions$^{19)}$, although, the solar value
is
missed by many standard deviations.

\begin{figure}
\hspace{4truecm}
\epsfysize=20truecm\epsfbox{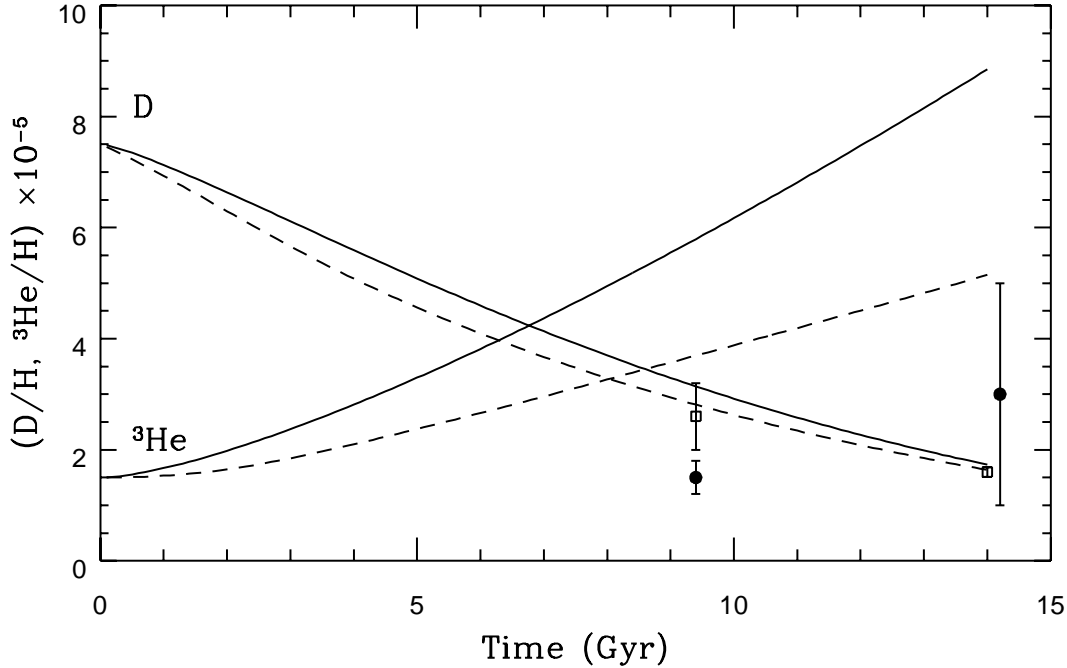}
\vspace{-10.5truecm}
\baselineskip=2ex
\caption { The evolution of D and \he3 with time.}
\end{figure}

\baselineskip=3ex
The overproduction of \he3 relative to the solar meteoritic value seems to be a
generic feature of chemical evolution models when \he3 production in low mass
stars is included. In ref. 14), a more extreme model of galactic chemical
evolution was tested.  There, it was assumed that the initial mass function
was time dependent in such a way so as to favor massive stars early on (during
the first two Gyr of the galaxy).  Massive stars are preferential from the
point of view of destroying \he3.  However, massive stars are also proficient
producers of heavy elements and in order to keep the metallicity of the disk
down to acceptable levels, supernovae driven outflow was also included.
The degree of outflow was limited roughly by the observed metallicity in the
intergalactic gas in clusters of galaxies. One further assumption was
necessary;
we allowed the massive stars to lose their \he3 depleted hydrogen envelopes
prior
to explosion.  Thus only the heavier elements were expulsed from the galaxy.
With all of these (semi-defensible) assumptions, \he3 was still overproduced
at the solar epoch as shown by the dashed curve in Figure 1. Though there
certainly is an improvement in the evolution of \he3, without reducing the
yields of low mass stars, it is hard to envision much further reduction in
the solar \he3 predicted by these models. The
only conclusion that we can make at this point is that there is clearly
something
wrong with our understanding of
\he3 in terms of either chemical evolution, stellar evolution or perhaps even
the
observational data.

Given the magnitude of the problems concerning \he3, it would seem unwise to
make any strong conclusion regarding big bang nucleosynthesis which is based on
\he3.  Perhaps as well some caution is deserved with regard to the recent D/H
measurements, although if the present trend continues and is verified in
several
different quasar absorption systems, then D/H will certainly become our best
measure for the baryon-to-photon ratio. Given the current situation however, it
makes sense to take a step back and perform an analysis of big bang
nucleosynthesis in terms of the element isotopes that are best understood,
namely, \he4 and \li7.

Monte Carlo techniques are proving to be a useful form of analysis regarding
big
bang nucleosynthesis$^{34,35)}$. In ref. 36), we performed just such an
analysis
using only \he4 and \li7. It should be noted that in principle, two elements
should be sufficient for constraining the one parameter ($\eta$) theory of BBN.
We begin by establishing likelihood functions for the theory and observations.
For example for \he4, the theoretical likelihood function takes the form
\beq
L_{\rm BBN}(Y,Y_{\rm BBN})
  = e^{-\left(Y-Y_{\rm BBN}\left(\eta\right)\right)^2/2\sigma_1^2}
\label{gau}
\eeq
where $Y_{\rm BBN}(\eta)$ is the central value for the \he4 mass fraction
produced in the big bang as predicted by the theory at a given value of $\eta$,
and $\sigma_1$ is the uncertainty in that  value derived from the Monte Carlo
calculations$^{35)}$ and is a measure of the theoretical uncertainty in the big
bang calculation. Similarly one can write down an expression for the
observational
likelihood function. In this case we have two sources of  errors as discussed
above, a statistical uncertainty, $\sigma_2$ and a systematic uncertainty,
$\sigma_{\rm sys}$.  Here, I will assume that the
systematic error is described by a top hat distribution$^{27,35)}$.
The convolution of the top hat distribution and the Gaussian (to describe
the statistical errors in the observations) results in the difference
of two error functions
\beq
L_{\rm O}(Y,Y_{\rm O}) =
{\rm erf}\left({Y - Y_{\rm O} + \sigma_{\rm sys}
     \over \sqrt{2} \sigma_2}\right) -
{\rm erf}\left({Y - Y_{\rm O} - \sigma_{\rm sys}
     \over \sqrt{2} \sigma_2}\right)
\label{erf}
\eeq
where in this case, $Y_{\rm O}$ is the observed
(or observationally determined)
value for the \he4 mass fraction. (Had I used a Gaussian to describe the
systematic uncertainty, the convolution of two Gaussians leads to a
Gaussian, and the likelihood function (\ref{erf}) would have taken a form
similar
to that in (\ref{gau}).

A total likelihood
function for each value of $\eta_{10}$ is derived by
convolving the theoretical
and observational distributions, which for \he4 is given by
\beq
{L^{^4{\rm He}}}_{\rm total}(\eta) =
\int dY L_{\rm BBN}\left(Y,Y_{\rm BBN}\left(\eta\right)\right)
L_{\rm O}(Y,Y_{\rm O})
\eeq
An analogous calculation is performed$^{36)}$ for \li7. The resulting
likelihood
functions from the observed abundances given in Eqs. (\ref{he4}) and (\ref{li})
is shown (unnormalized) in Figure 2.As one can see
there is very good agreement between \he4 and \li7 in vicinity
of $\eta_{10} \simeq 1.8$.

\begin{figure}
\hspace{0.5truecm}
\epsfysize=9truecm\epsfbox{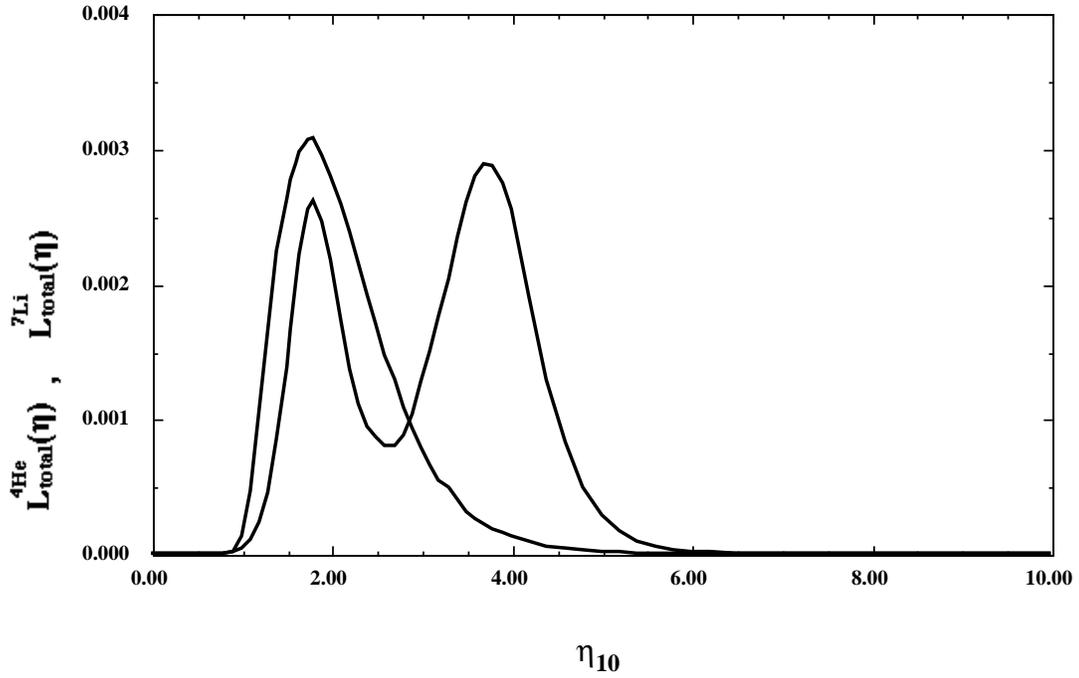}

\caption { \baselineskip=2ex Likelihood distribution for each of \he4 and \li7,
shown as a  function of $\eta$.  The one-peak structure of the \he4 curve
corresponds to its monotonic increase with $\eta$, while
the two-peaks for \li7 arise from its passing through a minimum.}
\label{fig:fig1}
\end{figure}

\begin{figure}
\hspace{0.5truecm}
\epsfysize=9truecm\epsfbox{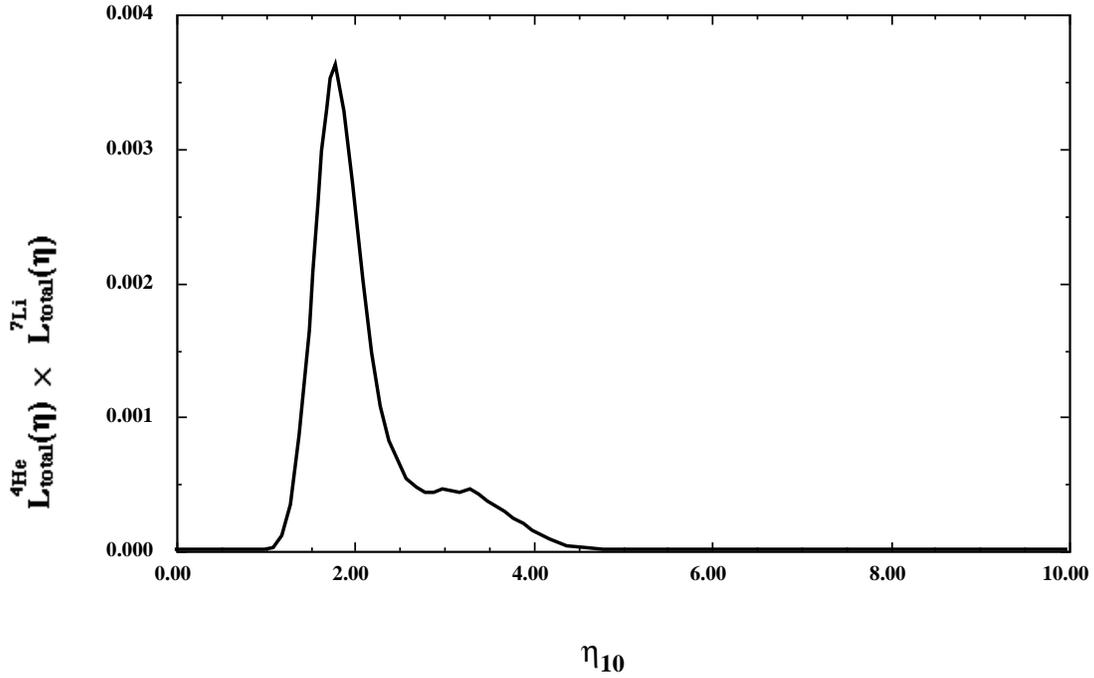}
\baselineskip=2ex
\caption { Combined likelihood for simultaneously fitting \he4 and \li7,
as a function of $\eta$.
}
\label{fig:fig2}
\end{figure}

\baselineskip=3ex

The combined likelihood, for fitting both elements simultaneously,
is given by the product of the two functions in Figure \ref{fig:fig1},
and is shown  in figure \ref{fig:fig2}.
{}From Figure \ref{fig:fig1} it is clear that \he4 overlaps
the lower (in $\eta$) \li7 peak, and so one expects that there will be
concordance,
in an allowed range of $\eta$ given by the overlap region.
This is what one finds in figure \ref{fig:fig2}, which does
show concordance, and gives an allowed (95\% CL) range of
$1.4 < \eta_{10} < 3.8$.  Note that
the likelihood functions shown in Figures \ref{fig:fig1}
and \ref{fig:fig2} are not normalized to unity.
An $\eta$ dependent normalization has however been included.
Any further normalization would
have no effect on the predicted range for $\eta$.

Thus,  we can conclude that
the abundances of
\he4 and \li7 are consistent, and select an $\eta_{10}$ range which
overlaps with (at the 95\% CL) the longstanding favorite
 range around $\eta_{10} = 3$.
Furthermore, by finding concordance
using only \he4 and \li7, we deduce that
if there is problem with BBN, it must arise from
D and \he3 and is thus tied to chemical evolution or the stellar evolution of
\he3. The most model-independent conclusion is that standard
BBN  with $N_\nu = 3$ is not in jeopardy,
but there may be problems with our
detailed understanding of D and particularly \he3
chemical evolution. It is interesting to
note that the central (and strongly)  peaked
value of $\eta_{10}$ determined from the combined \he4 and\li7 likelihoods
is at $\eta_{10} = 1.8$.  The corresponding value of D/H is 1.8 $\times
10^{-4}$ very close to the value  of D/H in quasar absorbers$^{16,18)}$.

\begin{figure}
\hspace{0.5truecm}
\epsfysize=9truecm\epsfbox{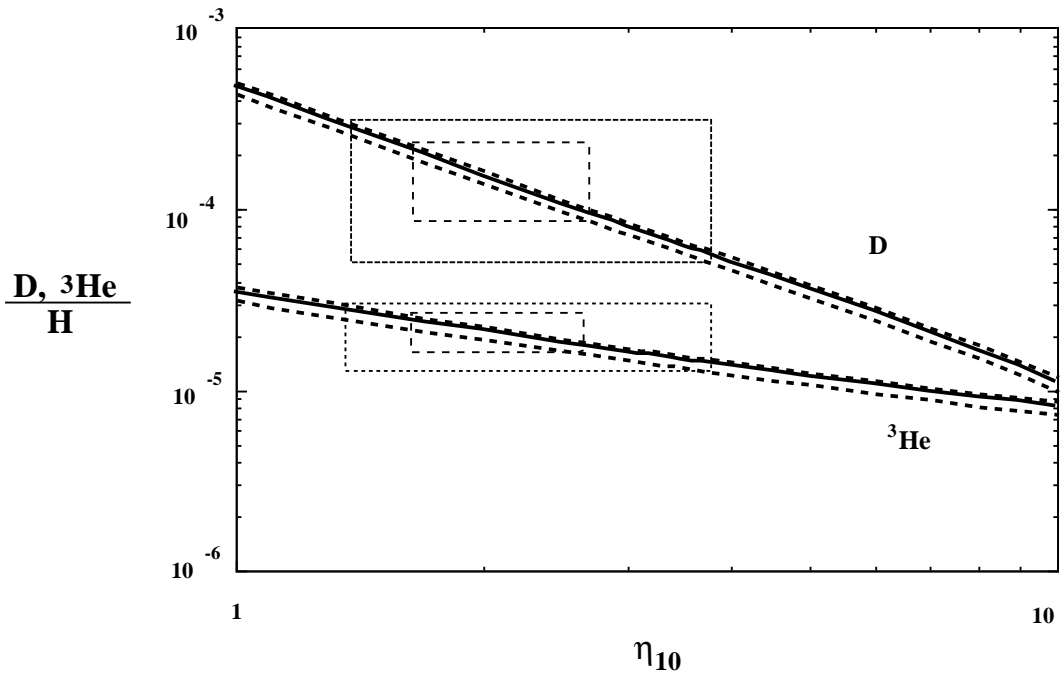}
{
\caption {\baselineskip=2ex  D/H and \he3/H as a function of $\eta_{10}$
from BBN along with the one $\sigma$ uncertainty from
Monte Carlo calculations$^{35)}$. Also shown are the values (demarcated
by rectangles) of
D/H and \he3/H consistent with
68\% (dashed) and 95\% CL (dotted)  likelihood values for $\eta_{10}$.
}}
\label{fig:dhe}
\end{figure}

\baselineskip=3ex
Since  D and \he3 are monotonic functions of $\eta$, a prediction for
$\eta$, based on \he4 and \li7, can be turned into a prediction for
D and \he3.  In Figure \ref{fig:dhe}, the abundances of
D and \he3 as a function of $\eta_{10}$ are shown along with the one $\sigma$
uncertainty in the calculations from the Monte Carlo results$^{35)}$.
The 68\% (dashed)  and 95\% CL (dotted) ranges for
D and \he3 as given by our likelihood analysis above are shown by a set of
rectangles. The corresponding 95\% CL ranges are D/H  $= (5.5 - 27)  \times
10^{-5}$ and  and \he3/H $= (1.4 - 2.7)  \times 10^{-5}$.

In summary, I would assert that one can only conclude that the present data on
the abundances of the light element isotopes is consistent with the standard
model
of big bang nucleosynthesis. Using the the isotopes with the best data, \he4
and
\li7, it is possible to constrain the theory, and obtain a best value for the
baryon-to-photon ratio of $\eta_{10} = 1.8$ with a 95\% CL range of 1.4 to 3.8.
This is a rather low value and corresponds to a baryon density $0.005 \le
\Omega_Bh^2 \le 0.014$, and would suggest that much of the galactic dark matter
is
non-baryonic$^{37)}$. These predictions are in addition consistent with recent
observations of D/H in quasar absorption systems. Difficulty remains however,
in
matching the solar \he3 abundance, suggesting a problem with our current
understanding of galactic chemical evolution or the stellar evolution of low
mass
stars as they pertain to \he3.

\vskip 1in
\vbox{
\noindent{ {\bf Acknowledgments} } \\
\noindent  I would like to thank M. Cass{\'e}, B. Fields, S. Scully,
D. Schramm, G. Steigman, J. Truran, and E. Vangioni-Flam for enjoyable
collaborations.  This
work was supported in part by DOE grant DE--FG02--94ER--40823.}


\end{document}